\documentclass[useAMS,usenatbib]{mn2e}
\usepackage{graphicx}
\DeclareGraphicsExtensions{.pdf, .jpg}
\usepackage{subfigure,lscape}
\newcommand{\integral}{{\it INTEGRAL}}
\newcommand{\rxte}{{\it RXTE}}
\newcommand{\xmm}{{\it XMM-Newton}}
\newcommand{\beppo}{{\it BeppoSAX}}
\newcommand{\mysou}{{GX~339$-$4}}
\newcommand{\cyg}{{Cyg~X$-$1}}
\def\ergcms{erg cm$^{-2}$ s$^{-1}$ }

\title[Spectral variability of \mysou]{Spectral variability of GX~339$-$4 in a hard-to-soft state transition
\thanks{Based on observations with {\it INTEGRAL}, an ESA  project with instruments and science data centre funded by ESA member states 
(especially the PI countries: Denmark, France, Germany, Italy,  Switzerland, Spain), Czech Republic and Poland, 
and with participation  of Russia and the USA.}}
\author[M. Del Santo et al.]{M. Del Santo$^{1}$, J. Malzac$^{2}$, E. Jourdain$^{2}$, T. Belloni$^{3}$, P. Ubertini$^{1}$\\
$^{1}$INAF/Istituto di Astrofisica Spaziale e Fisica Cosmica di Roma, via Fosso del Cavaliere 100, 00133 Roma, Italy \\
$^{2}$Centre d'Etude Spatiale des Rayonnements (CESR), OMP, UPS, CNRS; B.P. 44346, 31028 Toulouse Cedex 4, France \\
$^{3}$INAF/Osservatorio Astronomico di Brera, Via E. Bianchi 46, I-23807 Merate (LC), Italy}

\begin{document}

\date{Accepted 2008 July 02. Received 2008 June 26}

\pagerange{\pageref{firstpage}--\pageref{lastpage}} \pubyear{2008}

\maketitle

\label{firstpage}

\begin{abstract}
We report on \integral\ observations of the bright black-hole transient \mysou\ performed during the period August-September 2004,
including the fast transition (10 hr) observed simultaneously with \integral\ and \rxte\ on August 15th and previously reported.
Our data cover three different spectral states, namely Hard/Intermediate State,
Soft/Intermediate State and High/Soft State.
We investigate the spectral variability of the source across the different spectral states.
The hard X-ray spectrum becomes softer during the HIMS-to-SIMS transition,
but it hardens when reaching the HSS state. 
A principal component analysis demonstrates that most of the variability occurs through two independent modes:
a pivoting of the spectrum around 6 keV (responsible for 75\% of the variance) and an intensity variation of the hard component 
(responsible for 21\%). 
The pivoting is interpreted as due to changes in the soft cooling photon flux  
entering the corona, the second mode as fluctuations of the heating rate in the corona.
These results are very similar to those previously obtained for Cygnus X-1.
Our spectral analysis  of the spectra of \mysou\  shows a high energy excess with respect to pure thermal Comptonisation models in the HIMS: 
a non-thermal power-law component seems to be requested by data.
In all spectral states joint IBIS, SPI and JEM-X data are well represented by hybrid thermal/non-thermal Comptonisation ({\sc eqpair}). 
These fits allow us to track the evolution of each spectral components during the spectral transition. 
The spectral evolution seems to be predominantly driven by a reduction of  the ratio of 
the electron heating rate to the soft cooling photon flux in the corona, $l_h/l_s$.  
The inferred accretion disc soft thermal emission increases by about two orders
of magnitude, while the Comptonised luminosity decreases by at most a factor of 3.
This confirms that the softening we observed is due to a major increase in the flux of soft cooling photons 
in the corona associated with a modest reduction of the electron heating rate.

\end{abstract}


\section{Introduction}
Black hole candidates (BHCs) are known to show different spectral states in the X and $\gamma$-ray domain.
The two main spectral states are the Low/Hard State (LHS), with the high energy spectrum described 
by a cut-off power-law (typically $\Gamma$ $\sim$1.5 and $E_{cut} \sim$100 keV), and
the High Soft State (HSS) with a thermal component peaking at few keV and the 
high energy power-law much softer ($\Gamma > 2.2$) \cite{zdz00}.
Usually, this spectral variability is interpreted as due to changes in the geometry of 
the central parts of the accretion flow (see Done, Gierli\'nski \& Kubota 2007 for a recent review).
In the LHS the standard geometrically thin and optically thick disc \cite{ss73} would be truncated far away from the last stable orbit.
In the innermost parts, a hot accretion flow is responsible for the high energy emission,
via thermal Comptonisation of the soft photons coming from the truncated disc.  
In the HSS, the optically thick disc would extend close to the minimum stable orbit producing the dominant thermal component. 
The weak non-thermal emission at higher energy is believed to be due to up-scattering of the soft 
thermal disc emission in active coronal regions above the disc \cite{zdzg04}.

However, recently long \xmm\ observations of \mysou\ have shown that a standard 
thin accretion disc may remain at the innermost stable circular orbit around
the black hole also during the LHS (Miller et al. 2006, Reis et al. 2008)

Since its discovery \cite{markert73}, the X-ray binary \mysou\ has been thoroughly  studied at all wavelength from
radio to gamma-rays. Its star companion is still unknown, even though upper limit on the optical 
luminosity allowed to classify the source as Low Mass X-ray Binary (LMXB; Shahbaz, Fender, \& Charles 2001). 
Classified as BHC \cite{zdz98}, \mysou\ is a transient source 
spending long periods in outburst. 
Hynes et al. (2003) estimated a mass function of 
5.8 $\pm$ 0.5 $M_{\odot}$ and Zdziarski et al. (2004) derived a lower limit on the distance of 7 kpc.
Before the launch of \rxte\ the source had been
observed mostly in the Low/Hard spectral state (LHS), even though few spectral transitions to the Very High state (VHS)
were reported before \cite{miya91}. Thereafter \mysou\ remained bright and mostly in the LHS 
until 1999 when it went into 
quiescence. After the quiescent state observed by \beppo\ (Kong et al. 2000, Corbel et al. 2003),
\mysou\ showed two new outbursts: in 2002/2003 (Smith et al. 2002a,  Belloni et al. 2005) and   
in 2004 after one year in quiescence (Buxton et al. 2004, Belloni et al. 2004).  
The long-term variability of \mysou\ (1987-2004)
is extensively presented in Zdziarski et al. (2004). 

The spectral evolution of the outburst of a black hole in low-mass system is often investigated by plotting the flux 
of the source as a function of the X-ray hardness. 
In this Hardness Intensity Diagram (HID), \mysou\ usually follows a {\it q}-like pattern throughout the outburst 
(Homan \& Belloni 2005; Belloni 2005). 
The HSS and LHS correspond respectively to the left and right hand side vertical branches of the {\it q}. 
State transitions occur when the source crosses the upper or lower horizontal branches,
those constitute intermediate states between the LHS and HSS. 
Depending on the location of the source on a horizontal branch, Belloni et al. (2005) have identified two different flavours 
of intermediate state: the Hard Intermediate State (HIMS) and Soft Intermediate State (SIMS).
An alternative states classification defines the upper horizontal branch as VHS 
(see Tanaka \& Lewin 1995; van der Klis 1995; McClintock \& Remillard 2006).

During a typical outburst the source starts in a faint LHS, move upward along the LHS branch, then move leftward 
towards the HSS along the VHS branch, and when luminosity decreases the source moves down along the HSS before transiting 
right back to the LHS along the lower horizontal branch, and then back to quiescence in the LHS. 
A puzzling property of this evolution is that the spectral changes in \mysou\  lag
the variations of the luminosity: hard-to-soft state
transitions during the rising phase occur at higher luminosities than the soft-to-hard ones
during the declining phase. 
The so-called ``hysteresis'' \cite{zdzg04} is also observed 
in other LMXBs such as GRS 1758--258 and 1E 1740.7--2942 (Smith et al. 2002b; Del Santo et al. 2005).

The aim of this work is to use \integral\ \cite{winkler03} data collected during the softening of  the 2004 outburst to study  
the broad-band spectral evolution of \mysou\ during a hard-to-soft state transition. Some of these data obtained  
on 2004 August 14$^{th}$-16$^{th}$, right across the HIMS-SIMS  fast transition, are presented in Belloni et al. (2006). 
This transition, marked by the disappearing of type-C QPO and by the
appearance of a type-B QPO (Nespoli et al. 2003), is also associated to the ejection of fast relativistic jets,
which has led to the identification of a 'jet-line' in the HID (Fender, Belloni \& Gallo 2004). 
Moreover, focussing on the high energy part of the spectrum, Belloni et al. (2006) reported on  
the disappearing of the high energy cut-off (measured at 70 keV in the HIMS) in the SIMS.
Recently, a similar evolution of such cut-off
has been found for the BHC GRO J1655--40 \cite{joinet08}. These authors propose
that the cut-off increased significantly or vanished completely as the radio-jet turned off in the SIMS.  

Here, we extend the work preliminarily presented in Del Santo et al. (2006) by studying the spectral evolution of \mysou\ over 
a longer period of time covering almost the whole span of the LHS-HSS transition (from 2004 August 9 to 2004 September 11).
In this work, different Comptonisation models for spectral fitting have been used;
furthermore, a principal component analysis provided the variability modes of \mysou.

\section{Observations and data analysis}\label{sec:obsda}
\begin{figure}
\centering
\includegraphics[height=8.5cm,angle=+90]{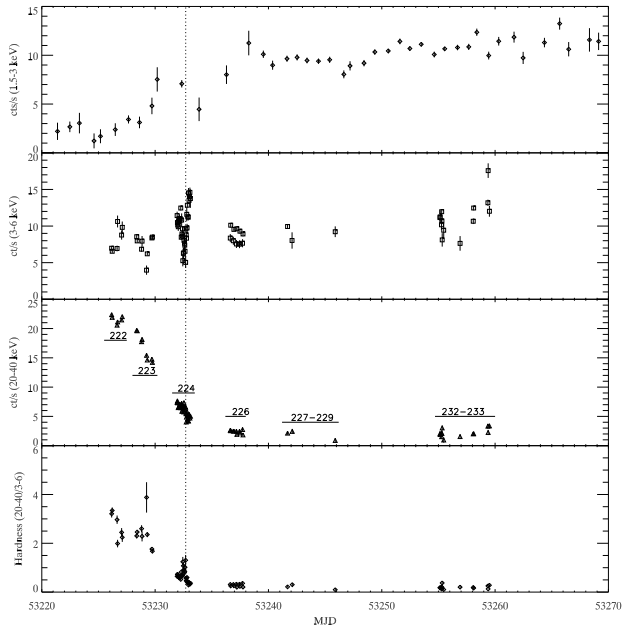}
\caption{From top to bottom: \mysou\ counts rate in  1.5-3 keV, 3-6 keV, 20-40 keV with ASM, JEM-X and IBIS/ISGRI, 
respectively. The related \integral\ orbits are also
marked on the IBIS light curve. IBIS/ISGRI to JEM-X hardness ratio is shown in the bottom panel.
The dotted line marks the fast transition HIMS-to-SIMS caught by Belloni et al. (2006). \label{fig:lc}}
\end{figure}

The 2004 X-ray activity of \mysou\ started on 
February 9$^{th}$ (Smith et al. 2004, Belloni et al. 2004).
This outburst was observed two times by the \integral\ satellite:
during the first part of the hard X-ray activity (19 February--20 March)
and during the decay of the hard X-ray flux, i. e. 9 August--11 September 
(the evolution of the flux and hardness of the source during this second period is shown in Fig.\ref{fig:lc}).
As expected in black hole transients \cite{hom-bel05}, 
at the beginning of the outburst \mysou\ was in the Low/Hard state (\integral\ 
observations presented in Joinet et al. 2007).
On 2004 August 15 (\integral\ orbit 224), a fast spectral state transition Hard-Intermediate (HIMS) 
to Soft-Intermediate (SIMS) was reported using simultaneous \rxte\ and \integral\ observations \cite{belloni06}.
Thereafter \mysou\ displayed a softer state as also shown by the hardness ratio (Fig.\ref{fig:lc}, bottom panel).

This paper is based on \integral\ public data collected during the second part of the outburst
(including the fast spectral state transition) between revolution 222 and 233. 
Because of the smaller JEM-X FOV ($4^{\circ} \times 4^{\circ}$) with respect to IBIS ($9^{\circ} \times 9^{\circ}$, fully coded),
we selected a data sub-set containing all common pointings of JEM-X \cite{lund03}, IBIS \cite{ubertini03} and SPI \cite{vedrenne03}, 
for a total of 76 Science Windows (SCWs\footnote{Each SCW corresponds to a pointing of about 30 minutes.}) (see Table \ref{tab:log} for details).

We reduced data of the IBIS low energy detector, ISGRI \cite{lebrun03}, and JEM-X data using the \integral\ 
Off-Line Scientific Analysis (OSA; Courvoisier et al. 2003). 

The SPI data analysis has been performed using a model fitting algorithm \cite{bouchet05}. With the sources emitting
in hard X-ray present within about 15 degrees around \mysou, we built a sky model consisting of 6 sources
(4U 1700--377, OAO 1657--415, 4U 1630--47, H 1636--536, IGR J16318--4848 and IGR J16320--4751),
considered as constant on over a revolution, except  4U 1700--377,
for which a $\sim$2000 seconds variability time-scale was used. The background pattern was built
from the revolution 220. 

The broad-band spectra were averaged in seven groups corresponding to the periods in Tab. \ref{tab:log}. 
A division of the different groups into spectral states based on timing analysis of \rxte\ data, i. e. on the type of the observed QPOs,
is also presented in the same table.

Because of the low statistics, SPI averaged spectra have been used only for spectral fitting of periods 1, 2 and 3. 
JEM-X1 camera spectra have been averaged in the energy range 4-20 keV for periods 1, 2, 3 and 4, and in the band 3-13 keV for 5 and 7. 
ISGRI and SPI data above 20 keV and 22 keV, respectively, have been used, while
the high energy limits were chosen between 200 and 600 keV depending on the statistics.  

Spectral fits were performed using the spectral X-ray analysis package XSPEC v. 11.3.1.

\section{Spectral variability}
\begin{figure}
%\centering
\includegraphics[height=8.5cm, angle=90]{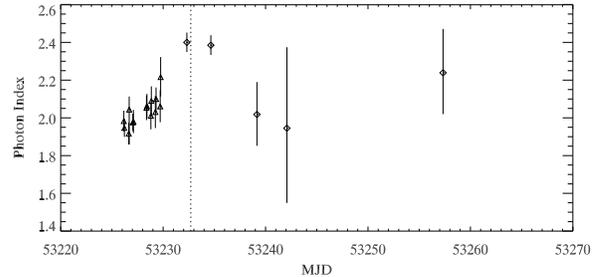}
\caption{\mysou\ photon index evolution.  The IBIS spectra collected scw by scw of periods 1 and 2 have been fitted with
a simple pawer-law in the range 20-80 keV (triangles).
Starting from period 3, all spectra have been averaged for each period and fitted with same model (diamonds). 
The HIMS-to-SIMS transition is marked (dotted line).}
\label{fig:gamma}
\end{figure}

As a first step in quantifying the spectral evolution of the source, we attempted to fit each of the 76 IBIS/ISGRI pointing 
spectra with a simple power-law ({\sc powerlaw} under XSPEC) in the energy band 20-80 keV. 
Due to the consistent flux decreasing of \mysou\ (Fig. \ref{fig:lc}, panel 3 from the top),
from period 3 onward the statistics by pointing is too poor to provide meaningful fit parameters. 
This is why from period 3 to 7, we fitted the spectra averaged over each data-set. 
The evolution of the photon index (Fig.~\ref{fig:gamma}) 
indicates that the spectrum softens when reaching the transition ($\Gamma$ goes from 1.9 to 2.4).
Thereafter it becomes harder when the soft state 
is reached ($\Gamma\sim 2$ in rev 226). 
The hard X-ray spectrum therefore appears softer around the end of the transition (i.e. in the SIMS) than in the HSS.

Then, we fitted the seven averaged broad-band spectra (IBIS, JEM-X and SPI when available) 
with a simple model combining a multicolor disc black-body ({\sc diskbb}),
a pawer-law (with a high energy cut-off when requested) and an emission line ({\sc gauss}) fixed at 6.4 keV.
Since the estimated absorption for \mysou\ is rather low and we did not have low-energy coverage,
we fixed it to 5$\times 10^{21}$ cm$^{-2}$ \cite{men-van97}.
Best-fit parameters of six periods (period 6 was avoided because of the statistic)
are summarised in Tab. \ref{tab:fit_sim}. 

The main spectral changes are the steepening of the power-law component and disappearance of the
high energy cut-off during the transition (periods 3 and 4).
This is in agreement with the results obtained using simultaneous \integral\ and \rxte\ 
observations and extensively discussed in Belloni et al. (2006).
During periods 5 and 7, the spectral parameters are consistent with being constant and the power-law is harder than in the SIMS (period 4). 
The disc temperature was frozen at 0.5 keV for periods 1 and 2, while for the others 
it was possible to constrain the parameter $kT_{bb}$.

\begin{figure}
\centering
\includegraphics[height=8cm, angle=+90]{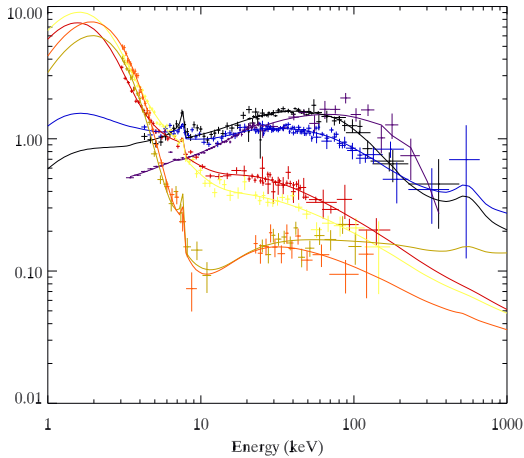}
\caption{Joint JEM-X, IBIS and SPI (only data-sets 1, 2 and 3 for the latter)
energy spectra of \mysou\ during periods 1 (black), 2 (blue), 3 (red), 4 (yellow), 5 (green), 7 (orange).
The data are fitted with the thermal/non-thermal hybrid Comptonisation
model {\sc eqpair} plus {\sc diskline} (see Tab. \ref{tab:fit}).
In order to give a comparison with a pure Low/Hard state spectrum,
we show the averaged spectrum collected during rev 175 (violet) and reported in Joinet et al. (2007).}
\label{fig:spectra}
\end{figure}

\section{Averaged spectra}
Since the cut-off power-law is simply an empirical model 
describing Comptonised spectra, we have used a further model which provides a physical description of X/$\gamma$-ray spectra.
In this section, we present spectral fits obtained using a hybrid thermal/non-thermal Comptonisation
model, namely {\sc eqpair} \cite{coppi99}, and relativistic iron line line emission and disc reflection. 
The best-fit parameters of six periods are shown in Tab. \ref{tab:fit}.

In the {\sc eqpair} model, emission of the disc/corona system is modeled by a
spherical hot plasma cloud with continuous acceleration of electrons illuminated by soft
photons emitted by the accretion disc. 
At high energies the distribution of electrons is non-thermal,
but at low energies a Maxwellian distribution with temperature $kT_{e}$ is established.

The properties of the plasma depend on its 
compactness $l = L {\it \sigma}_{T}/R{\it m_{e} c^{3}}$
where $L$ is the power of the source supplied by different components, $R$ the radius of the 
sphere ($10^{7}$ cm) and $\sigma_{T}$ is the Thomson cross-section. 
$l_{s}$, $l_{th}$, $l_{nth}$ and $l_{h}=l_{th}+l_{nth}$ correspond to the power in soft
disc photons entering the plasma, thermal electron heating, electron acceleration and total power supplied to the plasma.

The spectral shape is insensitive to the exact value of the compactness but it depends strongly on the the compactness ratios $l_h/l_s$ and $l_{nth}/l_h$. 
It is customary, when fitting data with this model, to fix $l_s$ to some reasonable value and parametrise the model with these compactness ratios.  
Leaving $l_{s}$ as free parameter we do not obtain any improvement on the $\chi^{2}$. We therefore fixed $l_{s} = 10$ following Gierli\'nski et al. (1999).  
On the contrary, we cannot fix $l_h$ because  it is not included as free parameter in the {\sc eqpair} version implemented in XSPEC.

If we knew with some accuracy the distance and size of the Comptonising region, the normalisation of the spectrum could provide 
a constraint on the  absolute value of the compactness. Since, however, there are large uncertainties on the size of the emitting region, 
and since there could be several active regions in the accretion flow contributing to the  observed flux, it is not straightforward 
to unambiguously constrain compactness from the observed luminosity. For this reason, in the version of {\sc eqpair} used to fit 
the data there is an additional normalisation parameter which actually disconnects $l_s$ (as a fit parameter) from the observed luminosity. 

The energy balance in the Comptonising medium depends mainly on the ratio $l_{h}/l_{s}$.  For larger values of $l_{h}/l_{s}$   the energy of the 
electrons is on average higher and, as a consequence, spectrum from Comptonisation is harder. 
From the best-fit parameters shown in Tab.~\ref{tab:fit}, we indeed measure lower values of  $l_{h}/l_{s}$ 
as the spectra become softer. Since in our spectral fits we chose to fix  $l_{s} = 10$, 
in this model the variations of $l_{h}/l_{s}$  are only due to changes in $l_{h}$.  

However, due to the free normalisation parameter of {\sc eqpair} we do not know if this what really happens in \mysou\  
(i.e . $l_s$ is constant while $l_{h}$ changes). We know for sure that we observe a change in $l_h/l_s$ 
but we cannot tell  from the fit parameters whether this evolution is  due to changes in the heating rate 
of the corona, changes in the luminosity of the disc, or both.

Nevertheless, from the best fitting model we can get important insights  
by computing the absolute disc and Comptonised fluxes (respectively $F_{bb}$ and $F_{Compt}$).
If  we consider the absolute values of $F_{Compt}$  and $F_{bb}$  (see Tab.~\ref{tab:fit}),  
we see that the thermal disc flux changes by more than one order of magnitude while the Comptonised flux decreases by a factor of about three. 
The softening we observe is therefore caused by a dramatic increase 
in the disc thermal flux  in the corona associated with a modest reduction of the electron heating rate. 

The size of the emitting region could be constant, or not. Assuming $l_s$ constant, as we did, it would imply that 
the size of the emitting region\footnote{or more exactly the number of X-ray emitting regions times their typical size} 
increases by a factor of $\simeq14$. On the contrary, the timing features usually indicate that the emitting region 
becomes smaller  when a source evolves from the LHS toward the  HSS (see e.g. Gilfanov, Churazov, Revnivtsev 1999). 
If this is the case in our observation,  the soft compactness actually increases during the transition. 
The bolometric luminosity increased by a factor of two during the whole spectral evolution.

The seed photons temperature $kT_{bb}$ was frozen at 300 eV for the spectra of revolutions 222 and 223; thereafter it was possible 
to constrain this parameter and observe a slight increase (Tab. \ref{tab:fit}) simultaneously with the flux of soft cooling.
This suggests that the increase of soft cooling photons is due to the higher temperature of the disc. 
   
However, the disc luminosity tends to decrease for larger temperatures.
The highest disc  temperatures, achieved during period 5 and 7, do not correspond with the strongest observed blackbody fluxes of period 3 and 4. 
This suggests that the disc emitting area and geometry of the corona are not constant during the transition. The radiating disc surface area appears smaller 
in the HSS (periods 5 and 7) than in the intermediate states (periods 3 and 4).  
This suggests that the corona becomes more compact as the disc temperature increases. 
This would be consistent with the truncation radius of the accretion disc moving closer to the black hole in the HSS. 

In order to consider the case of a hybrid plasma, the model allows for a fraction $l_{nth}/l_{h}$ of the power to be injected 
in the form of non-thermal electrons rather than heating of the thermalised distribution. The non-thermal electrons 
are injected with a power-law distribution $\gamma^{-G_{inj}}$, 
with Lorentz factors ranging from $\gamma_{min} = 1.3$ to $\gamma_{max} = 1000$.
From our spectral fits, we infer $G_{inj}$ values in the range 2--3, as expected form shock acceleration models.
We found large non-thermal fractions in all spectra, including the HIMS ones.
This is in agreement with previous {\sc eqpair} fits of Cygnus X-1 in intermediate states (Gierlinski et al. 1999; Malzac et al. 2006). 

However, we tried to fix $l_{nth}/l_{h} = 0$ for all spectra.
It resulted that only for the period 1 we obtained a good $\chi^{2}_{\nu}$ (i.e. 0.98) with 
remaining parameters consistent within the errors to the ones in Tab. \ref{tab:fit}.
For the other spectra, not negligible $l_{nth}/l_{h}$ fraction is required. 
Unfortunately the uncertainties on $l_{nth}/l_{h}$ are quite large and do not allow us to draw any conclusion regarding its evolution 
during the spectral transition. 

Another interesting free parameter is the Thomson optical depth ($\tau_{es}$). It is related to the ionization electrons only. 
The total optical depth ($\tau_{t}$) is the sum of the optical depth of
$e^{+}e^{-}$ pairs plus $\tau_{es}$.  In agreement with previous studies \cite{gier99} our fits show that the electron-positron 
pairs constitute only a small, if not negligible, fraction of the Comptonising leptons. An interesting trend is that the 
Thomson optical depth of the corona decreases significantly during the transition. This could be linked to the 'compactification' 
of the corona inferred from the evolution of the soft photon compactness and disc luminosity, if the electron density
remains constant while the size of the corona decreases.  

The disc reflection component is calculated for neutral material with standard abundances. 
We assumed an inclination angle of 50 $\deg$. This  component is then convolved with a general relativistic kernel assuming 
reflection on an accretion disc extending from 6 to 1000 gravitational radii around a Schwarzschild  black hole.  

We checked the need for the reflection component freezing $\Omega$/$2\pi$ at 0. 
The results show that the $\chi^2$ are consistently worst (F-test probabilities $<$ 10$^{-4}$), 
confirming that introducing the reflection component improves significantly the fits.
We can conclude that Compton reflection is requested by the data.
Our fits suggest that the reflection component increases up to  $\Omega$/$2\pi$=1 as the spectrum goes in softer states.
This is expected in the truncated disc model as the system evolves from a geometry where the reflecting disc 
is truncated at a large distance from the black hole (and therefore intercepts a small 
fraction of the coronal radiation) to a situation where the accretion disc is sandwiched by the illuminating corona. 
In period, 4, 5 and 7 the statistics was too low to constrain the reflection parameter and we decided to freeze it to $R=1$.

We used the XSPEC model {\sc diskline} to model the iron line \cite{fabian89}. 
Due to the energy resolution and line sensitivity,
the iron line study cannot be performed with JEM-X. The line energy 
was therefore  imposed at 6.4 keV. The inclination angle, inner radius of the disc and the disc emissivity law 
were fixed at the same values as in the Compton reflection model. 
The normalisation is therefore the only free parameter for the iron line emission. 

Spectra and models of the six different periods are shown in Fig. \ref{fig:spectra}.
For a comparison between all the states, the LHS spectrum observed in July \cite{joinet07}
has been plotted in the same figure.

\begin{figure}
\centering
\includegraphics[height=7cm,angle=-90]{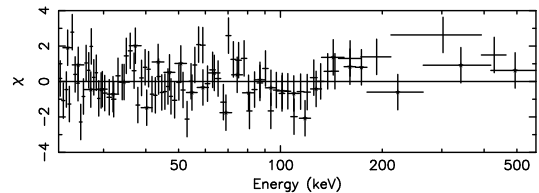}
\caption{Residuals obtained with a simple thermal Comptonisation model ({\sc compps}) 
by fitting IBIS and SPI spectra of rev 223. \label{fig:res}}
\end{figure}

\begin{figure}
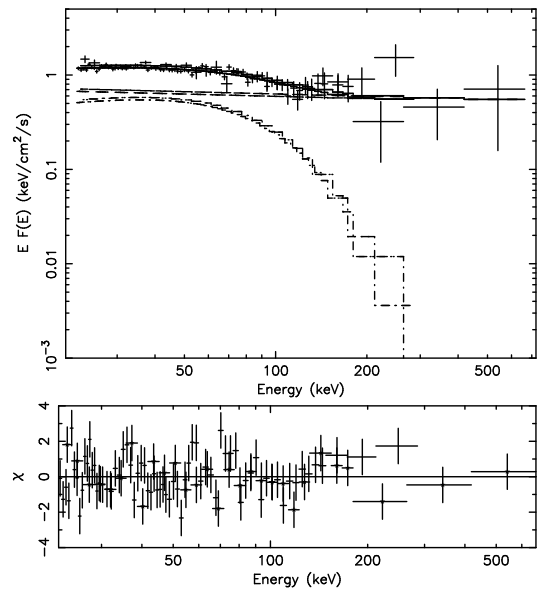

\centering
\includegraphics[height=7cm,angle=-90]{delsanto08_fig5.ps}
\includegraphics[height=7cm,angle=-90]{delsanto08_fig6.ps}
\caption{IBIS and SPI energy spectra and residuals of rev 223 fitted with a Comptonisation model 
({\sc compps}) plus power-law. 
\label{fig:he_spec223}}
\end{figure}

\begin{figure*}
\centering
\includegraphics[height=9.cm, width=4.6cm]{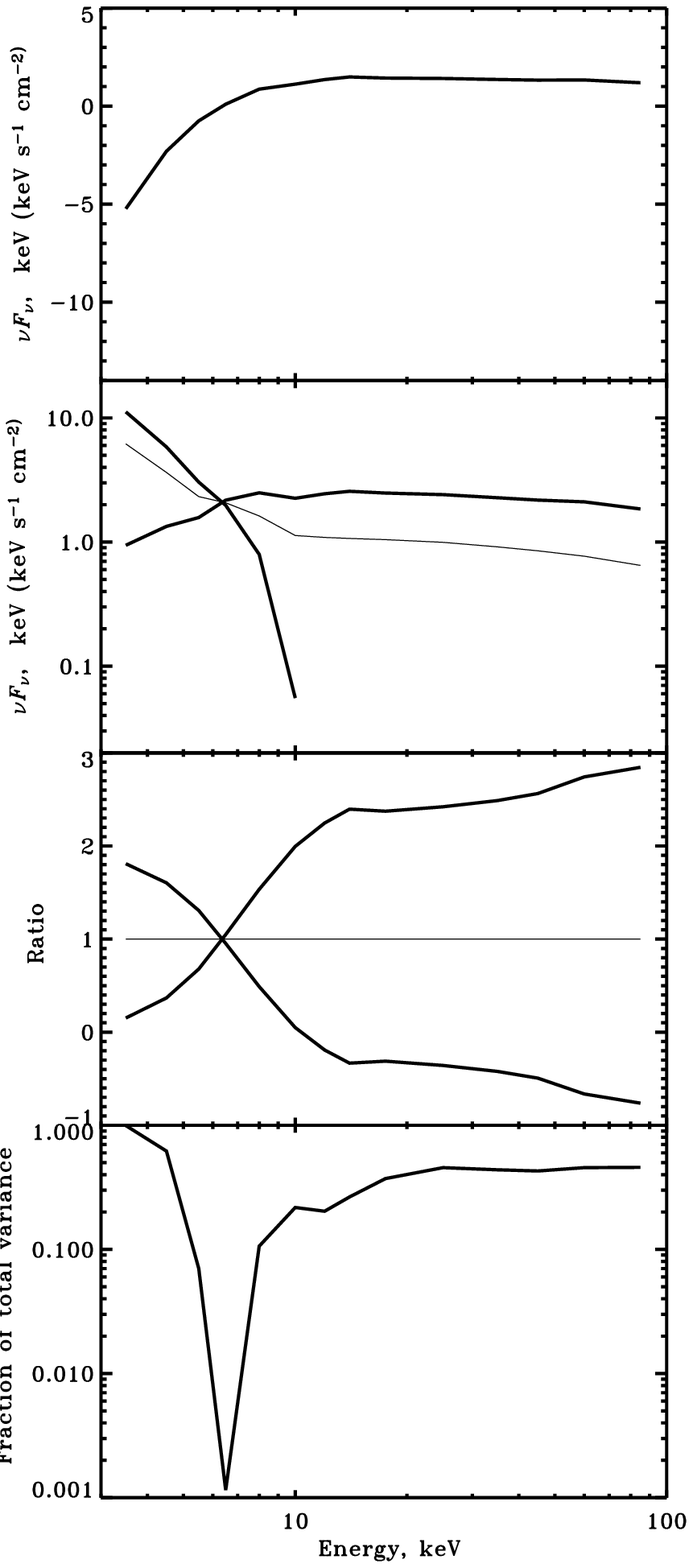}
\includegraphics[height=9.cm, width=4.6cm]{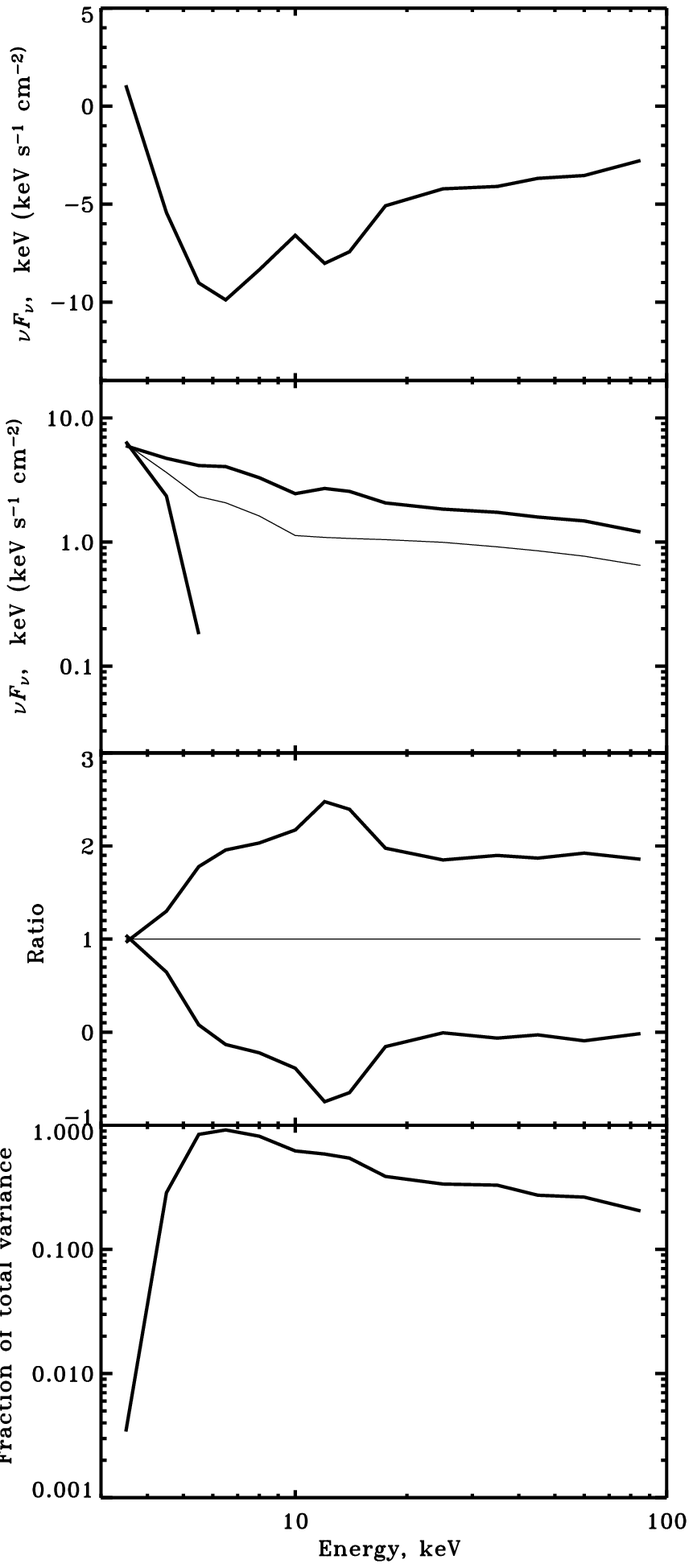}
\caption{PC1 ({\it left}) and PC2 ({\it right}). From top to bottom panels:  (i) the shape of the 
principal component (which would be added or subtracted to the average spectrum in order to reproduce the variability); 
(ii) the effects on the shape and normalisation of the spectrum  (time-averaged spectrum, shown by the light line,
and spectra obtained for the maximum and minimum amplitude of the component); (iii) the ratios of the
maximum (and minimum) spectrum to average; (iv) the contribution of the component to
the total sample variance as a function of energy. \label{fig:pca}}
\end{figure*}

\section{The high energy excess}

During the Low/Hard state of the \mysou\ outburst occurred in 1991, a high energy excess (above 200 keV) 
was observed with OSSE (Johnson et al. 1993; Wardzi{\'n}ski et al. 2002). 
In March 2004, i. e. during the increasing phase of the 2004 outburst, Joinet et al (2007) observed a similar feature with SPI
when \mysou\ was in its canonical LH state.
Similar features have been recently reported for other black hole transients 
in LH states observed during the decreasing phase of the outbursts (Kalemci et al. 2006; Kalemci et al. 2005).  
It is also more commonly associated with state transitions \cite{malzac06} and  with soft states \cite{gier99} (
in which case the non-thermal emission actually dominates over thermal Comptonisation).

Such a hard tail is usually attributed to the presence of a small fraction of non-thermal electrons in the hot Comptonising plasma.  
In the LHS and HIMS such a high energy  excess  could also be the result of  spatial/temporal variations 
in the plasma parameters (see e.g. Malzac \& Jourdain 2000).
In the HIMS of our observations, i.e. rev 222, 223 and 224.1, \mysou\ shows a highly significant evidence of a cut-off
around 70-100 keV. This indicates that soft photons are Comptonised by a thermal electrons population with a temperature $kT_{e}$ 
that can be measured by the cut-off energy. 
In addition, we found non-negligible values of $l_{nth}/l_{h}$  for revolutions 223 and 224.1,
indicating that a non-thermal emission is also requested by the data. 

In order to investigate furtherly the possible presence of a non-thermal component during the HIMS, 
IBIS and SPI spectra of the three HIMS periods (rev 222, 223 and 224.1) have been fitted
with the {\sc compps} model of Poutanen \& Svensson (1996) (see Tab.~\ref{tab:hefits}). 
Using this thermal Comptonisation model, some residuals are present at high energy, especially in the spectrum of rev 223 (Fig.~\ref{fig:res}). 
In order to mimic the presence of a non-thermal component, we added a power-law to the pure thermal Comptonisation model (Fig. \ref{fig:he_spec223}). 
This led to an improvement of the $\chi^2$ that is highly significant for orbit 223 (Tab.~\ref{tab:hefits}). 
The F-test probability that this improvement was by chance is  
$p_{{\sc compps}} = 6.6 \times 10^{-2}$.
On the contrary, because of the low statistics in rev 224.1, adding a power-law to the Comptonisation model 
is not required.   

{\sc compps} is a highly accurate iterative scattering Comptonisation model
in which subsequent photon scatterings are directly followed.
In practice this method is limited to Comptonisation in a plasma of reasonably small Thomson optical depth 
($\tau < 3$ for spherical geometry).  
In order to be sure that the presence of a non-thermal component does not depend 
on the specific Comptonisation model, we also fitted the data with the {\sc comptt} model  
of Titarchuk (1994) which is based on an approximate solution of the kinetic equation with some relativistic corrections. 
We performed the same fits as with {\sc compps}. As expected, we found  significantly different numerical values of the best-fit parameters  
(see Tab. {\ref{tab:hefits}}). However, we obtained a stronger indication regarding the presence of the non-thermal component: 
the addition of a power-law leads to a highly significant improvement of the $\chi^2$ both in spectra of rev 222 and rev 223.

\section{The principal Component Analysis}
In order to study the overall variability of \mysou\, we present in this section a further analysis, i. e. Principal
Component Analysis (PCA), based on a different approach.
The PCA is a powerful tool for multivariate data analysis used for a broad range 
of applications in natural as well as social science \cite{kendall80}. 
The main use of the PCA is to reduce the dimensionality of a data-set while keeping
as much informations as possible.
PCA transforms a number of (possibly) correlated variable in a smaller number of uncorrelated variables
called principal components. 

Details on the method, applied to the analysis of the spectral variability of 
Cyg X--1 with \integral, are given in Malzac et al. (2006). 
It provides us an approximation of the broad-band energy spectrum for each pointing
that will be used for a physical interpretation of the variability.
This method is much more convenient than fitting the spectra for each SCW,
also considering the large number of model parameters and the poor statistic of spectra in short (3-4 ks) exposure times.

Here, we applied a similar procedure to the state transition data of \mysou. 
We used {\it p}=76 spectra (one for each SCW) binned into {\it n}=15 bins corresponding to
 the following energy bands: 3-4, 4-5, 5-6, 6-7, 7-9, 9-11,
11-13, 13-15, 15-20, 20-30, 30-40, 40-50, 50-70, 70-100, 100-200 keV. 

This PCA showed that only two independent components contribute significantly to the variability of 
\mysou\ during the state transition. These two components are displayed in Fig.~\ref{fig:pca}.
The first  component  (PC1), responsible for 75\% of the sample variance, consists in a pivoting of the spectrum around 6 keV.  
The second component  (PC2) consists in intensity variation of the hard power-law component (almost constant slope)  
on top of constant soft component. It contributes to 21\% of the sample variance.
The timescale of the two variability modes is not firmly determined because our observations are not continuous.
We can only state that, it is comprised between 2 ks, i. e. the minimum duration of a pointing,
and one month, i. e. duration of the whole observation.

These results are remarkably similar to those obtained by  Malzac et al. (2006) during an intermediate state of \cyg. 
These authors found that the spectral variability occurred 
through two independent modes: the first consisted in changes in the overall luminosity on time scale 
of hours with almost constant spectrum (flaring mode); 
the second mode consisted in a pivoting of the spectrum around 10 keV. 
The flaring mode was interpreted as fluctuations of the heating rate in the corona possibly associated  to some erratic magnetic activity. 
The pivoting was associated to a mini-state transition. It was interpreted as due to changes in the soft cooling photon flux  
entering the corona as a consequence of variations of the temperature and luminosity of the optically thick disc.

Unlike \cyg, in \mysou\  the pivoting mode is more important than the flaring mode,  
probably  because \cyg~ was observed during an incomplete spectral transition hard-to-intermediate, 
while \mysou~ achieved finally a softer state. The presence of a pivot at 6 keV dominating the variability is apparent 
in the averaged spectra shown in Fig.~\ref{fig:spectra}. 
In the framework of the {\sc eqpair} model, this pivoting 
is clearly associated to the disc getting hotter, with the thermal component becoming prominent while the ratio $l_{\rm h}/l_{\rm s}$ 
was decreasing, leading to a softer and weaker hard X-ray spectrum.  
The interpretation of the pivoting mode in both \cyg\ and \mysou\ as due to variations in the 
soft cooling photon flux entering the corona is very plausible.

\section{Conclusions}

We have presented \integral~ observations of \mysou\ which covered part of the 2004 outburst. 
We followed the source spectral evolution through the Hard-Intermediate and Soft-Intermediate states, 
until the High-Soft state. 

During the HIMS-to-SIMS transition we found that the high energy cut-off disappears, 
in agreement with Belloni et al. (2006).
Furthermore, the hard X-ray spectrum appears softer during the transition states than in the HSS.

Our detailed spectral analysis of time averaged spectra at different stages of the transition confirms that 
the spectral transition is driven by changes in the soft cooling photon flux in the corona associated 
with an increase of disc temperatures (leading to dramatic increase of disc luminosity).
The measured disc temperature versus luminosity relation suggests that the internal disc radius decreases.
In contrasts, the heating rate of the electrons in the corona appears to remain nearly constant.  
Although other models such as dynamic accretion disc corona models cannot be ruled out, 
these results are consistent with the so-called truncated disc model (Done et al. 2007).

In all spectra, including those in of HIMS, we found a significant contribution from a non-thermal component. 
This component appears as a high energy excess above the pure thermal Comptonisation spectrum.  
We associated this component with the presence of a non-thermal tail in the distribution of the Comptonising electron.

Finally we analysed the variability of the source during the state transition with a PCA. 
The  PCA shows that the variability of \mysou~ can be described as produced through two independent modes: 
\begin{itemize}
\item{spectral evolution with the spectrum pivoting around 6 keV (PC1) contributing for 75\% of the observed variance. 
The pivoting mode may be interpreted as caused by changes in the soft cooling photons flux in the hot Comptonising plasma 
associated with an increase of temperature of the accretion disc. 
These changes are well in agreement with the variation in the flux of the soft cooling photons 
found by fitting averaged spectra with {\sc eqpair}.}
\item{intensity variation of the hard power-law component (almost constant slope)
on top of constant soft component (PC2) contributing at 21\%, also consistent with the variations of either the 
compactness ratio $l_{h}/l_{s}$ or the hard X-ray luminosity, both found by the spectral fitting. 
This variability mode could be due to magnetic flares or local instabilities in the corona.}
\end{itemize}

\section*{Acknowledgments}
Data analysis is supported by the Italian Space Agency (ASI),
via contract ASI/INTEGRAL I/008/07/0.
MDS thanks Memmo Federici for the \integral\ data archival support at IASF-Roma.
JM acknowledges support from CNRS.
TMB acknowledges support from the International Space Science Institute (ISSI)
and from ASI via contract I/088/06/0.

\begin{table*}
  \begin{center}
    \caption{Log of the \integral\ observations. A part the data-set 6 (because of the low exposure time),
all periods from 1 to 7 have been used for the averaged spectra analysis. The spectral states classification
is based on the timing analysis of \rxte\ data (Belloni et al. 2006; Belloni priv. comm.).}
    \vspace{1em}
    \renewcommand{\arraystretch}{1.5}
    \begin{tabular}{lrcccccc}
      \hline
      Period & Rev. & Start (MJD) &  End (MJD) & Exp. IBIS (ks) & Exp. JEM-X & SCW & State\\
      \hline
      1  & 222    & 53226.14 & 53227.12 & 24.8  & 24.3 & 6 & HIM \\
      2  & 223 & 53228.36 & 53229.78  & 32.1 & 31.8 &   8 & HIM \\
      3  & 224.1 & 53231.93  & 53232.65 & 48.5  & 48.6  & 22 & HIM \\
      4  & 224.2 & 53232.68 & 53233.13   & 30.4  & 29.0 & 14 & SIM\\
      5  & 226 & 53236.61 & 53237.77  & 33.3   & 32.5 & 10   & HS \\
      6  & 227-229 & 53241.67 & 53245.90 & 9.07 & 6.05 & 3   & HS \\
      7  & 232-233 & 53255.12 & 53259.51  & 25.2   & 25.4 & 13 & HS\\
      \hline \\
      \end{tabular}\\
 \end{center}
\label{tab:log}
 \end{table*}

\begin{table*}
  \begin{center}
    \caption{Spectral fitting parameters of joint IBIS, JEM-X and SPI (only data-set 1, 2 and 3 for the latter)
by using a simple model, namely a multicolor-disc black-body plus a power-law with a high energy cut-off (when requested).
In periods 1 and 2 the disc temperature has been frozen at 0.5 keV.}
    \vspace{1em}
    \renewcommand{\arraystretch}{1.5}
    \begin{tabular}{lrcccc}
      \hline
      Period & Rev. & $kT_{bb}$ (keV) &$\Gamma$   & $E_{c}$ (keV)& $\chi^2_{\nu}$(dof) \\
      \hline
      1  & 222    & (0.5) &$1.4 \pm 0.1$ & $71^{+8}_{-2} $ & 1.0(243) \\
      2  & 223 & (0.5)& $1.8 \pm 0.1$ & $117^{+43}_{-11}$ & 1.1 (214)\\
      3  & 224.1 & $0.70 \pm 0.03$  &   $2.1 \pm 0.1$ & $115^{+27}_{-23}$ & 1.2(216)\\
      4  & 224.2 & $0.69 \pm 0.03$ & $2.4 \pm 0.1$   & $-$ & 1.1(173)\\
      5  & 226 & $ 0.64 \pm 0.02$ &$1.8 \pm 0.2$ & $-$ & 1.1(121)\\
      7  & 232-233 & $0.64 \pm 0.02$  &  $2.0 \pm 0.1$ & $-$ & 1.0(98)\\
      \hline \\
      \end{tabular}
    \label{tab:fit_sim}
  \end{center}
\end{table*}

\begin{landscape}

\begin{table}
 \begin{center}
\caption{Best-fit parameters of the joint IBIS/ISGRI, JEM-X and SPI spectra (only rev 222, 223 and 224.1 for the spectrometer).
Fits have been performed simultaneously with {\sc eqpair} combined with {\sc diskline}. 
See text for the parameters description.}\label{tab:fit}
\renewcommand{\arraystretch}{1.3}
\begin{tabular}{cccccccccccccc} 
\hline
\hline
Period & Rev   &  $\rm l_h/l_s$ & $\rm l_{nth}/l_{h}$&$\tau _{es}$& kT$_{bb}$ &$\Omega$/${2\pi}$& $G_{inj}$&$\tau _{tot}$ &kT$_{e}$&$\chi^2_{\nu}$(dof) & 
\multicolumn{3}{c}{\rm Flux $\times 10^{-9}$} \\	
      &        &                   &            &        &     [eV]             &         &             &  &   [keV]      &   & \multicolumn{3}{c}{[\ergcms]} \\
      &        &                   &            &        &                      &         &             &  &              &   & $Bol$ & $bb$& $Compt$  \\
\hline
1 & 222&$4.36^{+1.02}_{-0.39}$ & $0.75^{+0.25}_{-0.14}$ & $2.58^{+0.24}_{-0.93}$ & (300) & $0.23^{+0.26}_{-0.13}$ & $2.75^{+0.46}_{-0.64}$ & 2.72 & 23.2 & 0.99(243)&
11.3 & 1.0 & 9.4 \\   
2& 223&$2.10^{+0.40}_{-0.39}$ &$1.0^{+0}_{-0.17}$  & $1.96^{+0.31}_{-0.49}$  & (300) & $0.6 \pm 0.3$   & $2.62^{+0.27}_{-0.37}$  & 2.05 & 19.2  & 1.17(219)&
12.4  & 2.1 & 8.9 \\
3&224.1&$0.20^{+0.05}_{-0.07}$ & $0.71^{+0.18}_{-0.09}$ & $0.18^{+0.12}_{-0.03}$ & $380^{+54}_{-57}$& $1.0^{+0.10}_{-0.35}$& $2.91^{+0.14}_{-0.22}$ &0.20 & 45.0 & 1.14(215)&
21.0  & 15.4 & 5.6 \\
4&224.2&$0.17^{+0.03}_{-0.02}$ & $0.56^{+0.11}_{-0.10}$ &$0.36^{+0.34}_{-0.12}$&  $388^{+76}_{-83}$ & (1) & $2.86^{+0.37}_{-0.18}$ &0.36 & 29.4 & 1.06(173)&
 24.3 & 17.8   & 6.5  \\
5&226&$0.15^{+0.13}_{-0.05} $ & $0.94^{+0.06}_{-0.10} $ &  $< 0.1$     & $495^{+27}_{-43}$  &  (1) &$1.9 \pm 0.3 $  & 0.1 & 21.7 & 1.15(121)&
13.8 & 10.9  & 2.9  \\
7&232-233&$0.07^{+0.04}_{-0.01}$ &$1.0^{+0}_{-0.2} $  &  $ < 0.7$ & $478^{+66}_{-83}$ & (1) & $2.8 \pm +0.5$   & 0.7 & 5.5 &  1.0(98)&
 18.4  & 13.8  & 4.2  \\
\hline
\end{tabular}
\vspace*{3 cm} 
\end{center}
\end{table}

\end{landscape}

\begin{table*}
\renewcommand{\arraystretch}{1.5} 
\begin{center}
\caption{Parameters of the {\it SPI} and {\it IBIS} spectra
obtained by fitting data (in 20-500 keV energy range) with thermal-Comptonisation models and thermal-Comptonisation plus power-law.
{\sc compps} and {\sc comptt} have been used as thermal-Comptonisation models.
Comptonisation temperature ($kT$) and Thomson depth ($\tau$) are free parameters;
$\Gamma$ is the power-law photon index.
In the {\sc compps} model the black body temperature 
of the soft seed photons was fixed at 0.1 keV in all fits. Reflection component (parameter $\Omega$/$2\pi$) is only 
present in {\sc compps}.}       
\label{tab:hefits}      
\begin{tabular}{l c c c c c c c c} 
   
\hline\hline
Rev & Model    & $kT$  (keV)                     & $\tau$        &  {\rm tau-y}   &  $\Omega$/$2\pi$             &  $\Gamma$         & $\chi^2_{\nu}$(dof) & F-test $p$\\ 
\hline                    
222 & & & & & & & &\\
& {\sc comptt}       & $27 \pm 2$           & $1.7 \pm 0.1$        &      $-$  &         $-$        &   $-$           & $1.2(111)$ & $-$\\
& {\sc comptt+po}    & $21 \pm 3$       &  $2.1^{+0.6}_{-0.3}$ &   $-$    &        $-$          &  $1.8^{+0.3}_{-0.7}$&  $1.1(109)$ & $9.2 \times 10^{-4}$\\
&  {\sc compps}      & $42^{+4}_{-3}$ &  $-$        &  $3.1 \pm 0.2$& $0.6 \pm 0.3$     &  $-$ &  $1.08(104)$  & $-$\\
& {\sc compps+po}   & $27^{+10}_{-5}$      &  $-$   &  $5.2^{+4.8}_{-1.6}$   & $ 0.4^{+0.6}_{-0.4}$ & $ 2.0^{+0.2}_{-0.1}$      & $1.05(102)$ & 0.1 \\
\hline                    
223& & & & & & & &\\    
& {\sc comptt}       & $56^{+31}_{-17} $ & $0.6 \pm 0.3$ & $-$    &    $-$            &  $-$          & $1.1(104)$ & $-$\\
& {\sc comptt+po}    & $17^{+8}_{-4}$     &  $2.6^{+6.1}_{-0.6}$ &   $-$  &  $-$    &    $2.2^{+0.8}_{-0.9}$  & $1.0(102)$ &  $3.4 \times 10^{-3}$ \\
&  {\sc compps}      & $86^{+10}_{-6}$ &    $-$  &      $1.1 \pm 0.1$ &   $ 0.7^{+0.9}_{-0.4}$       &  $-$             &   $1.2(103)$ & $-$ \\
&  {\sc compps+po}   &  $20^{+31}_{-4}$  & $-$  &   $6.8^{+0.5}_{-4.9}$  & $0.2^{+2.4}_{-0.1}$    &  $2.08^{+0.06}_{-0.09} $   & $1.1(101)$ &  $6.6 \times 10^{-2}$ \\
\hline\hline                                
\end{tabular}

\end{center}
\end{table*}

\end{document}